# Giant nonlocal edge conduction in the axion insulator state of $MnBi_2Te_4$

Yaoxin Li[a,1], Chang Liu[a,b,c,d,1], Yongchao Wang[a,1], Zichen Lian[a,] Shuai Li[e,f,g,h], Hao Li[i,j], Yang Wu[j,k], Hai-Zhou Lu[e,f,g,h,*], Jinsong Zhang[a,m,*] , Yayu Wang[a,l,m*]

[a] *State Key Laboratory of Low Dimensional Quantum Physics, Department of Physics, Tsinghua University, Beijing 100084, China*

[b] *Beijing Academy of Quantum Information Sciences, Beijing 100193, China*

[c] *Beijing Key Laboratory of Opto-electronic Functional Materials & Micro-Nano Devices, Department of Physics, Renmin University of China, Beijing 100872, China*

[d] *Key Laboratory of Quantum State Construction and Manipulation (Ministry of Education), Renmin University of China, Beijing, 100872, China*

[e] *Shenzhen Institute for Quantum Science and Engineering and Department of Physics, Southern University of Science and Technology, Shenzhen 518055, China*

[f] *Shenzhen Key Laboratory of Quantum Science and Engineering, Shenzhen 518055, China*

[g] *Quantum Science Center of Guangdong-Hong Kong-Macao Greater Bay Area (Guangdong), Shenzhen 518045, China*

[h] *International Quantum Academy, Shenzhen 518048, China*

[i] *School of Materials Science and Engineering, Tsinghua University, Beijing 100084, China*

[j] *Tsinghua-Foxconn Nanotechnology Research Center, Department of Physics, Tsinghua University, Beijing 100084, China*

[k] *College of Science, Beijing University of Chemical Technology, Beijing 100029, China*

[l] *New Cornerstone Science Laboratory, Frontier Science Center for Quantum Information, Beijing 100084, China*

[m] *Hefei National Laboratory, Hefei 230088, China*

* Corresponding author.

*Email address:* luhz@sustech.edu.cn (H.-Z. Lu), jinsongzhang@tsinghua.edu.cn (J. Zhang), yayuwang@tsinghua.edu.cn (Y. Wang).

[1] These authors contributed equally to this work.

## ABSTRACT

The recently discovered antiferromagnetic (AFM) topological insulator (TI) $MnBi_2Te_4$ represents a versatile material platform for exploring exotic topological quantum phenomena in nanoscale devices. It has been proposed that even-septuple-layer (even-SL) $MnBi_2Te_4$ can host helical hinge current with unique nonlocal behavior, but experimental confirmation is still lacking. In this work, we report transport studies of exfoliated $MnBi_2Te_4$ flakes with varied thicknesses down to the few-nanometer regime. We observe giant nonlocal transport signals in even-SL devices when the system is in the axion insulator state but vanishingly small nonlocal signal in the odd-SL devices at the same magnetic field range. In conjunction with theoretical calculations, we demonstrate that the nonlocal transport is via the helical edge currents mainly distributed at the hinges between the side and top/bottom surfaces. The helical edge currents in the axion insulator state may find unique applications in topological quantum devices.

## 1. Introduction

In the field of topological quantum materials, searching for electronic phases with nontrivial band topology and novel transport phenomena in the two-dimensional (2D) limit is a perpetual quest [1,2]. Magnetic topological insulator (TI) with gapped Dirac cone and broken parity-symmetry serves as an ideal platform for exploring exotic topological quantum phenomena [1–3]. The surface of magnetic TI is expected to manifest a half-quantized Hall conductivity $\sigma_{xy} = e^2/2h$, where $h$ is Planck's constant and $e$ is electron charge [4–11]. This phenomenon, known as the parity anomaly, plays significant roles in many condensed matter systems [12–15]. One representative example is the quantum anomalous Hall (QAH) effect realized in ultrathin ferromagnetic TI films with parallel magnetizations on the top and bottom surfaces [16–21]. The chiral edge state forms along the side surface, contributing a quantized Hall conductance $e^2/h$ hence an integer Chern number $C = 1$ [1,2,4]. The axion insulator represents another exotic topological quantum phase that is related to the half-quantization phenomena [10,22,23]. The magnetoelectric responses of an insulator can be described by a quantized topological axion $\theta$ term, which takes $\pi$ for the axion insulator [4,8–10,24]. Therefore, the axion insulator hosts the quantized topological magnetoelectric effect with the coefficients related to the fine structure constant. To realize the axion insulator, the Dirac point of the top and bottom surfaces should be gapped by opposite magnetizations so that there is no bound state at the side surfaces [25–27]. An early idea to realize the axion insulator is to construct magnetic TI heterostructures with opposite magnetizations [25–27]. In the charge transport, the axion insulator is characterized by a zero Hall plateau and strongly reduced longitudinal conductivity, as has been observed in several experiments [25–31].

Despite the realization of axion insulator phase in different TI configurations, the nature of the zero Hall plateau state is still highly controversial. The observation of strongly reduced Hall and longitudinal conductivities cannot distinguish the axion insulator from a normal insulator [32]. There are distinctively different views regarding whether the axion insulator can host one-dimensional (1D) edge conduction channels, which are anticipated from the surface Hall current pictures [6,7,10,23,24,32,33]. One way to detect the edge conduction is through the nonlocal transport experiment, as has been performed in the $\nu = 0$ quantum Hall state of TI [34]. However, direct observation of nonlocal transport in the axion insulator state is still absent, despite the recent observation of half-quantized $\sigma_{xy}$ by local transport in semi-magnetic TIs [35].

The recently discovered $MnBi_2Te_4$ antiferromagnetic (AFM) TI provides an ideal platform for exploring the underlying edge transport. It has been proposed and demonstrated that even-septeple-layer (even-SL) $MnBi_2Te_4$ devices host the axion insulator state at the low-magnetic-field AFM state and Chern insulator state at the high-magnetic-field ferromagnetic (FM) state [28–30,36]. In contrast to conventional magnetic TI heterostructures, the $\theta$ term in $MnBi_2Te_4$ is protected by the combined symmetry $S = \Theta T_{1/2}$, where $\Theta$ is the time-reversal (TR) operator and $T_{1/2}$ is the half translation operator [6,37–39], hence is more stable against the TR-symmetry breaking by magnetic order. The suppression of the scattering events by magnetic impurities in exfoliated few-nanometer-thick $MnBi_2Te_4$ devices makes the edge transport more distinguishable than the doped TI systems.

Here, we report nonlocal transport studies on exfoliated $MnBi_2Te_4$ devices with even- and odd-SL thicknesses at varied magnetic fields, temperatures ($T$s) and gate voltages ($V_g$s). In even-SL $MnBi_2Te_4$, we observe giant nonlocal resistance ($R_{NL}$) in the axion insulator state, where the longitudinal resistance ($R_{xx}$) is unquantized but proportional to the channel length. In contrast, in odd-SL $MnBi_2Te_4$ devices the nonlocal signal becomes vanishingly small at the same magnetic field range, with pronounced anomalous Hall resistance. Theoretical calculations show that the electrical transport of $MnBi_2Te_4$ in the axion insulator state is via a pair of counterpropagating hinge currents, which arise from the opposite magnetizations of the top and bottom surfaces [4–10,22,23].

## 2. Materials and methods

By direct reaction of high-purity $Bi_2Te_3$ and MnTe with the ratio of 1:1, high-quality $MnBi_2Te_4$ single crystals are grown in a vacuum-sealed silica ampoule. The mixture is first heated to 973 K and then slowly cooled down to 864 K. The crystallization occurs during the prolonged annealing progress at this temperature. The phase and crystal quality are examined by X-ray diffraction on a PANalytical Empyrean diffractometer with Cu Kα radiation.

$MnBi_2Te_4$ flakes are exfoliated onto 285-nm-thick $SiO_2$/Si substrates by using Scotch tape method. The substrates are cleaned in air plasma for 3 min at ~125 Pa pressure before exfoliation. Thick flakes surrounding target sample are manually removed off by a sharp Be-Cu needle. The electrodes are patterned by using standard electron beam lithography, followed by thermal evaporation of Cr/Au. All the device fabrication processes are carried out in argon-filled glove box with the $O_2$ and $H_2O$ levels lower than 0.1 ppm. All the devices are covered by 400 nm thick poly(methyl methacrylate) to avoid contamination from the air during the fabrication processes and the transferring between glove box and cryostats.

Electrical transport measurements are performed in a cryostat with base temperature of 1.6

K and magnetic field up to 9 T. The Hall, local and nonlocal voltages are detected by using lock-in amplifiers with alternating-current (AC) of 200 nA generated by a Keithley 6221 current source. In the two-terminal measurements, an AC voltage bias of 10 mV is applied by a lock-in amplifier and the current is detected simultaneously. The frequencies of AC source current and source voltage are both 12.357 Hz. The $V_g$ was applied by a Keithley 2400 source meter.

To calculate the energy dispersion of $MnBi_2Te_4$ thin films, we first regularize the non-magnetic effective continuum model on a cubic lattice. The out-of-plane magnetism then added to each layer, with opposite directions in adjacent layers. The energy dispersion is calculated with periodic boundary condition along $x$-direction and open boundary condition along $y$-direction (6- or 7-SL-thickness in $z$-direction). To see the spatial distribution of states, after squaring the wave functions, the orbital and spin components are summed. To calculate the resistance, eight leads are attached to the system, same as the experimental setup. The transmission between leads are calculated based on the wave function approach implemented in Kwant software package. Then, the resistance is calculated by Landauer–Büttiker formula.

## 3. Results

The magnetic and crystal structure of $MnBi_2Te_4$ is shown in Fig. 1a. For an even-SL $MnBi_2Te_4$ device (such as 6 SLs), the magnetic moments between neighboring SLs are AFM-coupled at zero magnetic field. The opposite surface Hall conductance contributes to zero Hall plateau in charge transport. Fig. 1b shows the device structure and electrode configuration for our transport measurements. The photograph of a 6-SL $MnBi_2Te_4$ (Sample S4) is shown in Fig. 1c. To confirm its axion insulator behavior, we first calibrate its transport properties using the standard four-probe method. As shown in Fig. 1d, when $E_F$ is tuned within the bandgap, the 2D longitudinal resistivity ($\rho_{xx}$) shows an overall insulating behavior as lowering $T$s. The Néel temperature $T_N \sim 20$ K can be clearly identified from the $\rho_{xx}$ peak, consistent with previous reports [28–30,36]. Interestingly, the maximum $\rho_{xx}$ is only around 95 kΩ and shows a tendency of saturation at low $T$. Fig. 1e presents the magneto-resistivity (MR) and Hall resistivity ($\rho_{yx}$) measured at 1.5 K when $V_g = 21$ V. The MR maintains a high value ($\sim 4\ h/e^2$) at low magnetic fields. Meanwhile, the $\rho_{yx}$ forms a zero plateau in the field range $|\mu_0 H| < 3.5$ T. Fig. 1f shows the sheet conductivity data calculated by $\sigma_{xx} = \rho_{xx}/(\rho_{xx}^2 + \rho_{yx}^2)$ and $\sigma_{xy} = \rho_{yx}/(\rho_{xx}^2 + \rho_{yx}^2)$. The Hall conductivity $\sigma_{xy}$ has a zero plateau in AFM state as well. All these features are typical signatures of an axion insulator. As the magnetic field increases, there is a $\sigma_{xx}$ peak around 4.4 T which indicating the gap closing during the topological phase transition from the axion insulator to Chern insulator [28]. At 9 T, the system enters the Chern insulator state. The $\rho_{xx}$ drops to 0.002 $h/e^2$, accompanied by the increasing of $\rho_{yx}$ to a nearly quantized value of 0.997 $h/e^2$. Similarly, $\sigma_{xx}$ is vanishing small and $\sigma_{xy}$ presents the quantized plateau at $e^2/h$, suggesting the dissipationless chiral edge transport.

To reveal the nature of the transport of the axion insulator, we perform systematic nonlocal transport measurements on two 6-SL devices, with the schematic configurations shown in Fig. 2a and d. The resistance value $R_{mn,pq} = V_{pq}/I_{mn}$ is obtained by the voltage difference between electrodes $p$ and $q$ divided by the current flowing from electrodes $m$ to $n$. Fig. 2b displays the variation of both local resistance ($R_{\mathrm{L}}$) and $R_{\mathrm{NL}}$ as a function of $V_{\mathrm{g}}$ at zero magnetic field. As $E_{\mathrm{F}}$ is tuned within the bandgap (10 V $\leq V_{\mathrm{g}} \leq$ 25 V), the local resistance $R_{15,23}$ (black curve) exhibits a plateau-like behavior, suggestive of underlying conducting channels that saturate the total resistance [40,41]. When $E_{\mathrm{F}}$ is tuned away from the gap, the local resistance $R_{15,23}$ decreases gradually due to the contribution from bulk carriers.

Next, we focus on the nonlocal resistance $R_{37,ij}$ ($i$ and $j$ are the neighboring electrodes) with the current flowing between electrodes 3 and 7 as shown in Fig. 2a. By tuning $V_{\mathrm{g}}$ close to the charge neutral point (CNP), we find that all the nonlocal resistances, $R_{37,21}$, $R_{37,18}$, $R_{37,45}$, and $R_{37,56}$, display a peak larger than 50 kΩ, as shown in Fig. 2b. Remarkably, the maximum value of $R_{37,21}$ reaches the same level as local resistance $R_{15,23}$. Compared to the vanishing nonlocal signals in trivial bulk insulators (Fig. S1 online) and valley Hall systems [42–44], such a large $R_{\mathrm{NL}}$ provides strong evidence that the charge transport in the axion insulator state is carried by edge conductions. The little quantitative difference among four nonlocal resistances originates from different effective channel lengths and thus the current along those sample edges. The presence of edge channel can account for the plateau behavior of local resistance $R_{15,23}$, which is the parallel connection of conducting edge and insulating bulk near the CNP. When the bulk state becomes conductive as $E_{\mathrm{F}}$ lies in the conduction and valence bands, the current mainly flows through the bulk and the nonlocal resistance $R_{37,21}$ drops quickly to as low as 1.7% and 3.4% of the local resistance $R_{15,23}$ at $V_g = 0$ and 50 V respectively (see Fig. S2 online for more detailed data). In Fig. 2c, we plot the field dependence of $R_L$ and $R_{NL}$ at $V_g = 21$ V, the CNP. All the nonlocal curves show similar behaviors as the local resistance $R_{15,23}$. In the axion insulator regime ($|\mu_0 H| < 3.5$ T), $R_{NL}$ increases rapidly with magnetic field, which probably arises from the strong suppression of bulk conduction. In the Chern insulator regime at high fields, the dissipationless chiral edge state leads to a vanishing $R_{NL}$ as demonstrated in the magnetically doped TIs [18,45].

In Fig. 2d–f, we present the nonlocal data in a different electrode configuration with the current flowing between two neighboring electrodes 4 and 5. All nonlocal resistance $R_{45,32}$, $R_{45,21}$, $R_{45,18}$, and $R_{45,87}$ exhibit consistent behaviors as $R_{37,ij}$. Interestingly, we find that the maximum value of $R_{45,kl}$ ranges from 12 to 20 kΩ, which is approximately 1/4 of the extrema of $R_{37,ij}$ (50 – 90 kΩ). In fact, such behavior is close to the expectation of Landauer–Büttiker formalism for helical edge conduction in an eight-terminal device [46], in which $R_{15,23} = R_{37,21} = 4R_{45,21}$ (see Section C of the Supplementary materials for details). Similar behaviors have been demonstrated in the well-known quantum spin Hall (QSH) state of HgTe quantum wells with helical edge states [40,41]. Here, the observed resistance relation, $R_{15,23} \approx R_{37,ij} \approx 4R_{45,kl}$, provides strong support for the existence of the counterpropagating edge currents in the axion insulator state [7,24,32,33,47] and excludes other origin of nonlocal transport [31,42–44,48–51] (see section C of Supplementary materials). Consistent results are confirmed in another 6-SL $MnBi_2Te_4$ Sample S6 as shown in Fig. S6 (online).

Fig. 3a shows the variation of nonlocal resistances $R_{45,kl}$ with $T$ at $V_g = 21$ V, where the values decrease gradually with increasing temperature to $T \sim 16$ K due to the contribution from bulk conduction by thermal activation. Above that, the values of $R_{45,kl}$ rise unexpectedly till $T \sim 20$ K, and then decrease again at higher $T$s. The rising nonlocal signal around $T_N$ can be explained by the recovery of TR-symmetry in the paramagnetic phase, which turns the 6-SL $MnBi_2Te_4$ into a QSH insulator with helical edge states dominating transport [37]. In order to further exclude other origins of nonlocal transport, the scaling relation between the local resistance $R_{15,23}$ and the nonlocal resistance $R_{45,32}$ at varied $T$s is presented in Fig. 3b. In spin or valley Hall systems, the nonlocal transport is present in bulk conducting channel and give rise to the cubic relation $R_{\mathrm{NL}} \propto R_{\mathrm{L}}^3$ according to the diffusion mechanism [42–44,51]. The strong deviation from cubic relation at low $T$s shown in Fig. 3b can exclude these diffusion origins (see more in Section C of the Supplementary materials).

To separate the bulk and edge conductions, we use a specific electrode configuration as illustrated in the insets of Fig. 3c. Following previous studies [36,52], we apply a small bias voltage between electrode 3 and 7, and measure the current flowing through electrode 7 with electrode 1 and 5 grounded. Under this configuration, the measured conductance is mainly contributed by the bulk because the currents through edges are all shorted and flow to ground through electrode 1 and 5 before entering the current meter. As shown in Fig. 3c, the total resistance $R_{37,37}$ shows a clear plateau-like feature when $E_{\mathrm{F}}$ is close to the CNP, consistent with the local resistance $R_{15,23}$ (Fig. 2b and e). In contrast, the bulk is highly insulating with the maximum resistance exceeding 100 MΩ, two orders of magnitude higher than $R_{37,37}$, which provides strong evidence for the dominance of edge conduction. To further demonstrate the 1D nature of transport, we perform systematic transport measurements in nine 6-SL Hall bar devices with different length and width, and the device parameters are listed in Table S1 (online). The length dependence of $R_{xx}$ in the axion insulator state is displayed in Fig. 3d, and $\rho_{xx}$ is displayed in Fig. S6d (online). The linear dependence of $R_{xx}$ on channel length, regardless of the Hall bar width, and random distribution of $\rho_{xx}$ undoubtedly prove that the conduction is determined by 1D edge rather than surface or bulk states. With decreasing channel length down to 2 μm, $R_{xx}$ approaches the quantum resistance $h/e^2$ (dashed magenta line) expected for the half-quantized helical edge transport in the Hall bar configuration. So far we have not observed the quantization behavior like that in the QSH effect [41], and it remains debatable whether the edge transport in axion insulator can be quantized or not. As has been discussed in theory, the edge current is not a topologically protected mode like the chiral edge state in the QAH effect [53]. Therefore, the scattering between counterpropagating currents [54] with reducing thickness and the short dephasing length may destroy the quantization [55–58].

To further demonstrate the close relation between the edge transport and axion insulator state, we perform similar transport measurements on 7-, 8- and 9-SL devices. Unlike the 6- or 8-SL devices, the top and bottom surfaces of odd-SL devices have the same magnetization in the low-field AFM state. As shown in Figs. 4a–d, giant nonlocal signals are observed for both 6- and 8-SL devices when $V_g = V_g^0$, i.e., $E_F$ is tuned to the CNP. However, for odd-SL devices, the maximum nonlocal resistances are merely 0.12% and 0.47% of the local resistance for 7- and 9-SL respectively. Such low nonlocal signals can be attributed to the unidirectional chiral edge currents on the side surfaces. Fig. 4e shows the summarized $R_{NL}/R_L$ values of different samples. The even-odd oscillations clearly demonstrate that the edge conduction in the axion insulator state is tied with the opposite magnetizations of the top and bottom surfaces. When all the magnetic moments are polarized along the same direction at high magnetic fields, the quantized Chern insulator state forms and the chiral edge states lead to vanishing $R_{NL}$ regardless of thickness, as shown in Fig. S8c, f and I (online).

The nonlocal transport and the edge conduction strongly indicate that the charge transport in the axion insulator of even-SL $MnBi_2Te_4$ is carried by a pair of counter-propagating currents, reminiscent of the QSH effect [40,41,56,58,59] and other nontrivial zero Hall states at high magnetic fields [34,47,54,60–62]. In these previous reports, the helical edge modes are true bound states at the sample boundaries of a pure 2D system and exhibit quantized transport characteristics. In the axion insulator, however, the 1D helical-like currents are the collective modes of all bulk states according to theories [7,32,33,53,63]. Furthermore, the well-separated top and bottom surfaces [33,64–66], and the interlayer AFM order along *c*-axis, make it likely to be defined in a 3D system, rather than QSH insulator in a pure 2D system.

To gain more theoretical insights on the nature of edge transport and even-odd oscillation of nonlocal resistance, we calculate the energy dispersion in the axion insulator state of 6-SL $MnBi_2Te_4$ as displayed in Fig. 4f. The color indicates the spatial distribution of edge currents in *y*-direction by subtracting the components in minus-*y*-direction from plus-*y*-direction, where the components in *z*-direction are summed. Therefore, the red and blue points mark the states distributed at the two opposite edges along *x*-direction. Considering the out-of-plane magnetic order in $MnBi_2Te_4$, we then focus on the distribution of these states along *z*-direction. The chemical potential is set to slightly deviate from zero for the convenience of calculations [32,33,53]. As shown in Fig. 4g, the two-fold degenerate states with different momentum, marked by purple and green squares respectively in Fig. 4f, concentrate on the different diagonal hinges within *y*–*z* plane. These four conducting channels constitute the spatially separated helical-like currents and the chirality originates from the gapped top and bottom surfaces due to parity anomaly [7,33,53]. In contrast, the edge conduction in 7-SL $MnBi_2Te_4$ device exhibits chiral nature with symmetric distribution on the side surface due to the presence of in-gap chiral edge state and the same chirality of top and bottom hinge currents (Fig. S9a and b online), although zero-field $\rho_{yx}$ is not quantized in our odd-SL samples with thickness larger than five SLs. This scenario is also supported by the calculated local and nonlocal resistances in 6- and 7-SL $MnBi_2Te_4$ shown in Fig. S9c–f (online), which qualitatively agrees with the experimental results.

## 4. Discussion and conclusion

Finally, we discuss other possibilities for the edge conduction accounting for the observed nonlocal transport. Naively, trivial factors such as the dangling bond state [67], gate-induced edge charge accumulation [68], and gapless side surface [69,70] are all possible to give rise to trivial edge state. However, these factors are incompatible with the dependence of nonlocal transport on magnetic field and film thickness in our experiments, and thus can be largely excluded. Firstly, trivial edge states are resistive in micrometer scale and nearly insensitive to magnetic fields, and thus can survive at high magnetic fields [69]. Apparently, it is contrary to the quantized $\rho_{yx}$ and vanishing $\rho_{xx}$ for the Chern insulator as shown in Fig. 1e and Fig. S8 (online). Secondly, the contribution from trivial edge state should be more pronounced in thicker samples. In our experiments, however, the nonlocal signals have even-odd oscillation properties. Thirdly, trivial edge states are not protected by any symmetry or topology, whose contribution to transport would be strongly suppressed by impurities and defects. The small mobility (~ 112 $cm^2/(V\ s)$) of our $MnBi_2Te_4$ devices makes the trivial edge state transport negligibly small, if it exists at all.

Despite the good agreement between our nonlocal transport results and the helical current proposed in $MnBi_2Te_4$ axion insulator [33,70], we cannot completely exclude the possibility of diffusive helical edge states arising from random potential effect calculated by a 2D model [70]. In Section J of the Supplementary materials, we discuss the possibility of gap reducing by tuning the interlayer coupling. We find that the hinge-distribution feature of the edge conduction survives, which is consistent with previous calculations for AFM $MnBi_2Te_4$ [32,33]. To completely clarify this issue, there are proposals for future experiments, such as edge conduction in single-surface device [32], nonlocal transport in thinner samples [70] and the power-law decay of hinge current distribution [53].

In conclusion, the giant nonlocal resistance in even-SL $MnBi_2Te_4$ demonstrates the existence of 1D edge conduction channels that are consistent with the proposed helical hinge current picture. The hinge current is a new edge conduction mechanism owing to the nontrivial bulk band structure topology in the axion insulator state. It not only provides important insights on the unique transport properties of magnetic TIs, but also opens new opportunities for novel topological quantum device applications based on nano-flakes of $MnBi_2Te_4$.

## Conflict of interest

The authors declare that they have no conflict of interest.

## Acknowledgments

We thank Yang Feng and Xiaodong Zhou for helpful discussions. This work was supported by the Innovation Program for Quantum Science and Technology (2021ZD0302502), the National Key R&D Program of China (2018YFA0307100 and 2022YFA1403700) and the National Natural Science Foundation of China (NSFC; 51991340, 21975140, 11925402 and 12274453). Chang Liu was supported by Open Research Fund Program of the State Key Laboratory of Low-Dimensional Quantum Physics (KF202204). Hai-Zhou Lu was supported by the Guangdong Province (2020KCXTD001 and 2016ZT06D348), and Center for Computational Science and Engineering of SUSTech. Yayu Wang was supported by the New Cornerstone Science Foundation through the New Cornerstone Investigator Program and the XPLORER PRIZE.

## Author contributions

Yayu Wang and Jinsong Zhang supervised the research. Yaoxin Li, Chang Liu, Yongchao Wang and Zichen Lian fabricated the devices and performed the transport measurements. Hao Li and Yang Wu grew the $MnBi_2Te_4$ crystals. Shuai Li and Hai-Zhou Lu performed the theoretical calculation. Jinsong Zhang, Yayu Wang, Yaoxin Li and Chang Liu prepared the manuscript with comments from all authors.

## Figures

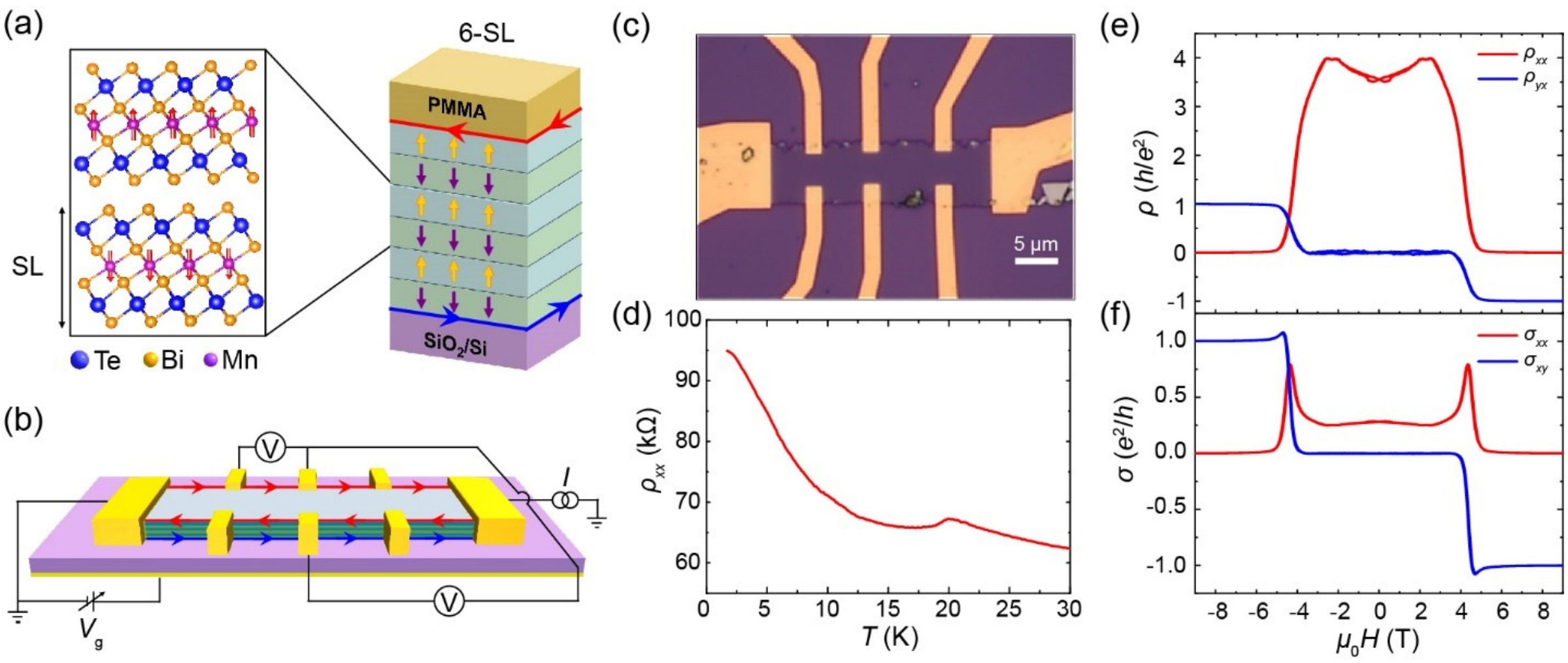

**Fig. 1.** Transport characterization of the 6-SL $MnBi_2Te_4$ device. (a) Crystal and magnetic structure of $MnBi_2Te_4$. The yellow/purple arrows in each SL indicate the magnetic moments of Mn atoms. The red/blue arrowed lines illustate the chiral hinge currents. (b) Measurement configuration of the 6-SL device. (c) Photograph of the 6-SL device. (d) Temperature dependence of $\rho_{xx}$ at $V_g$ = 21 V. (e), (f) Magnetic-field dependence of $\rho_{xx}$ and $\rho_{yx}$ (e), $\sigma_{xx}$ and $\sigma_{xy}$ (f) at $T$ = 1.5 K and $V_g$ = 21 V.

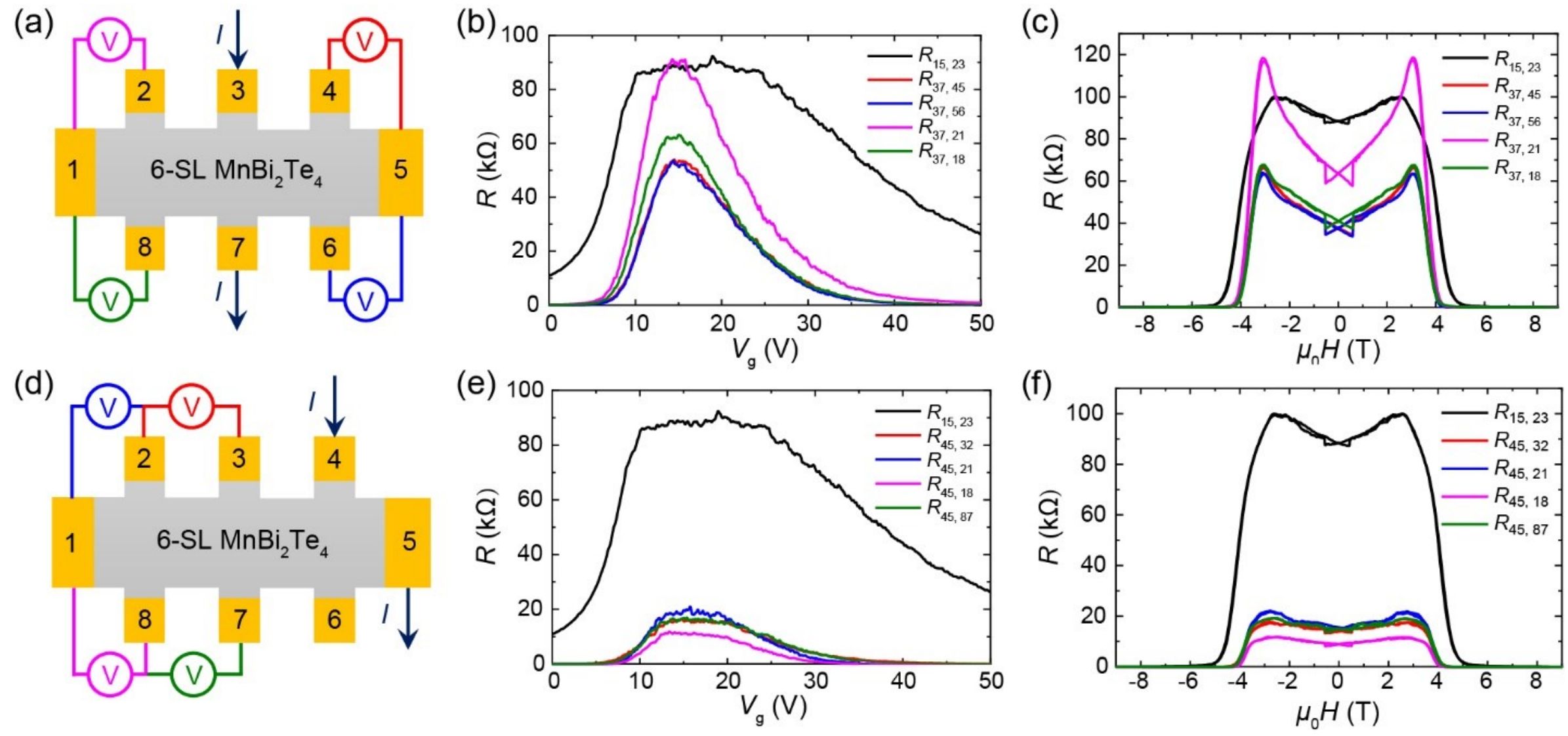

**Fig. 2.** Nonlocal transport in 6-SL $MnBi_2Te_4$ device at $T = 1.6$ K. (a) Schematic configuration of nonlocal transport measurements with the current between electrode 3 and 7. (b) $V_g$ dependent $R_L$ and $R_{NL}$ at $T = 1.6$ K and $\mu_0 H = 0$ T with the configuration shown in (a). (c) Magnetic-field dependence of $R_L$ and $R_{NL}$ with configuration shown in (a) at $T = 1.6$ K and $V_g = 21$ V. (d) Electrode configuration for nonlocal transport with the current between electrode 4 and 5. (e), (f) $V_g$ and magnetic-field dependence of $R_L$ and $R_{NL}$ with the configuration shown in (d).

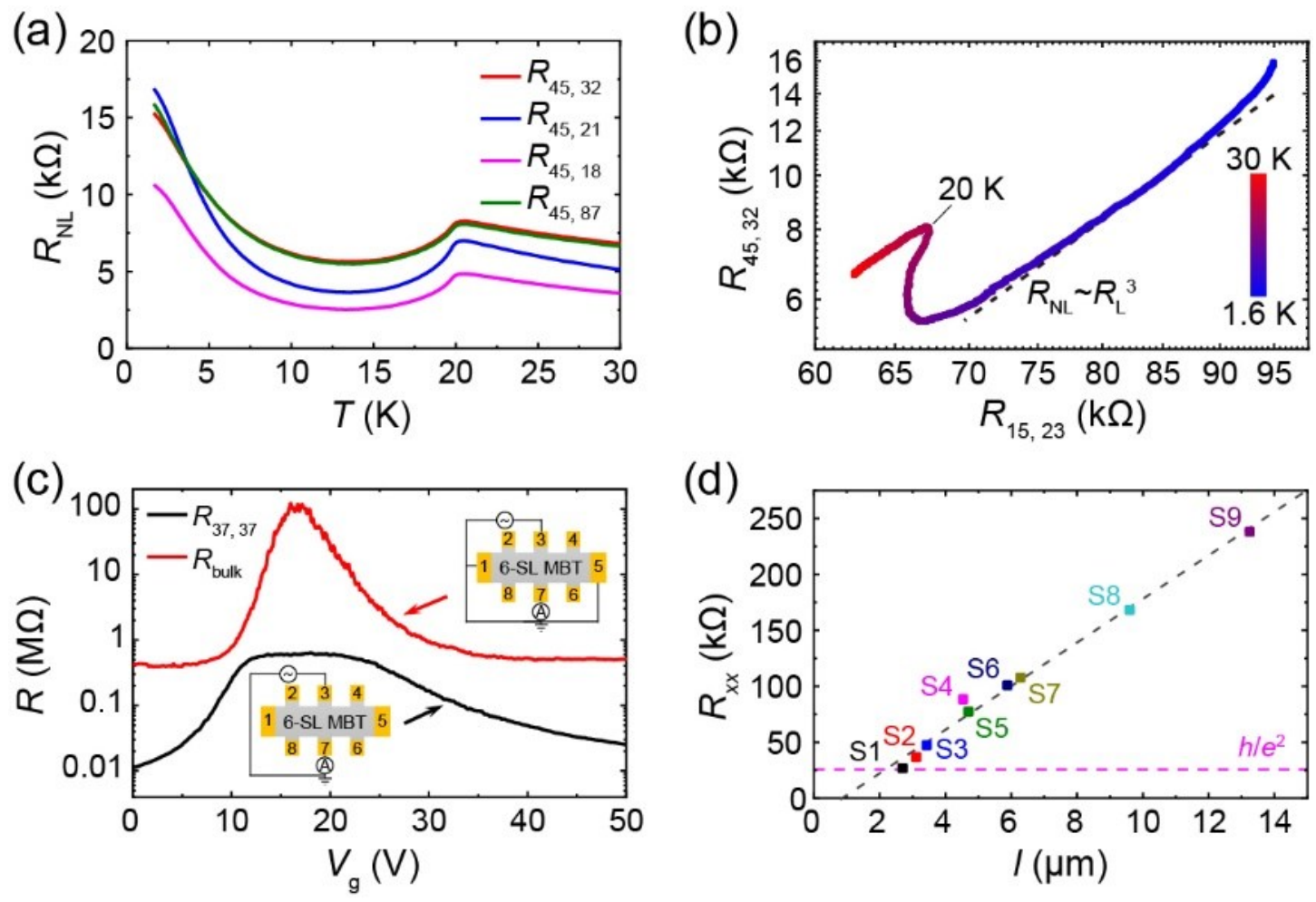

**Fig. 3.** Temperature evolution of nonlocal transport and edge conduction in the axion insulator state. (a) Temperature evolution of the $R_{NL}$s at $\mu_0 H = 0$ T and $V_g = 21$ V. (b) The scaling relation between local resistance $R_{15,23}$ and nonlocal resistance $R_{45,32}$ at varied $T$s with $V_g = 21$ V. (c) Gate dependence of two-terminal resistance $R_{37,37}$ and bulk resistance $R_{bulk}$ at $\mu_0 H = 0$ T and $T = 1.6$ K. The insets illustrate measurement configurations of $R_{bulk}$ (top) and $R_{37,37}$ (bottom). An AC voltage is applied at electrode 3 and the current flowing through electrode 7 is measured. (d) $R_{xx}$ values of nine 6-SL devices show a linear dependence on the channel length. The magenta dashed line marks quantized resistance.

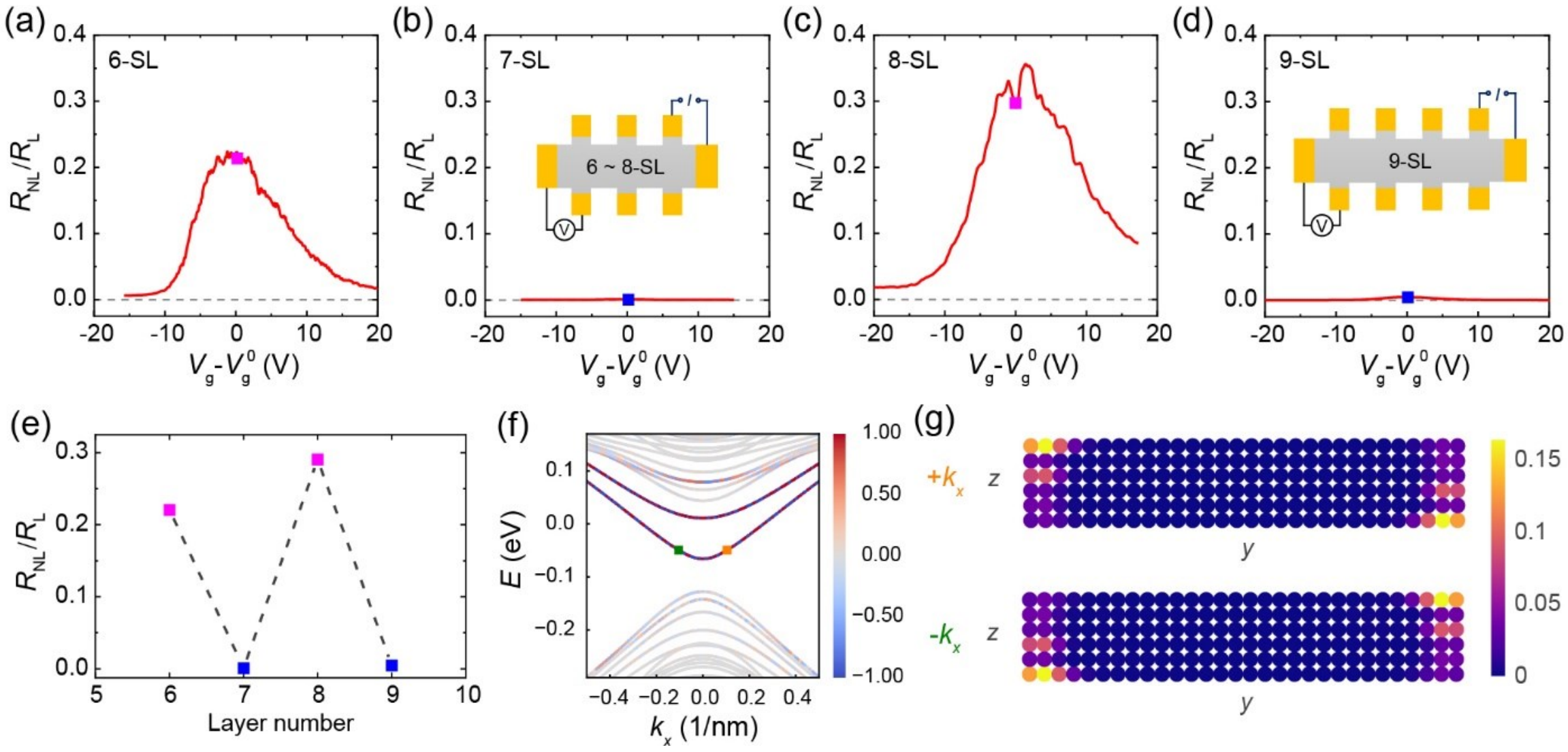

**Fig. 4.** Distinct edge current in axion insulator state of $MnBi_2Te_4$. (a)–(d) $V_g$ dependence of $R_{NL}/R_L$ in $MnBi_2Te_4$ at 6-, 7-, 8- and 9-SL devices. Inset of (b) shows the configurations of nonlocal measurement for 6-, 7- and 8-SL devices. Inset of (d) shows that for 9-SL device. (e) Thickness-dependent $R_{NL}/R_L$ when $V_g = V_g^0$. (f) Energy dispersion of 6-SL $MnBi_2Te_4$ and the state distribution in *y*-direction. (g) The spatial distribution of the states marked by orange ($+k_x$) and green ($-k_x$) squares in (f).